\documentstyle[12pt,psfig,epsf]{article}
\def\reference{\parskip 0pt\par\noindent\hangindent 0.5 truecm}

\def\deg   {$^\circ$}

\def\phiI  {$\varphi_i$}
\def\phiIV {$\varphi_4$}

\def\Em    {${\cal E}_m$}

\def\tauLL {\tau_{\scriptscriptstyle LL}}

\def\aB    {\alpha_{\scriptscriptstyle B}}

\def\Ha    {${\rm H}\alpha$}
\def\eg    {{\it e.g.,\ }}

\def\etal  {{\it\ et al.}}

\def\intensity{\ifmmode{{\rm erg\ cm}^{-2}{\rm\ s}^{-1}
      {\rm\ Hz}^{-1}{\rm\ sr}^{-1}}
      \else {erg cm$^{-2}$ s$^{-1}$ Hz$^{-1}$ sr$^{-1}$}\fi}
\def\flux{\ifmmode{{\rm erg\ cm}^{-2}{\rm\ s}^{-1}}\else {erg
cm$^{-2}$ s$^{-1}$}\fi}
\def\fluxdensity{\ifmmode{{\rm erg\ cm^{-2}\ s^{-1}\ Hz^{-1}}}\else {erg
cm$^{-2}$ s$^{-1}$ Hz$^{-1}$}\fi}
\def\phoflux{\ifmmode{{\rm photons\ cm}^{-2}{\rm\ s}^{-1}}\else {photons
cm$^{-2}$ s$^{-1}$}\fi}

\begin{document}

\title{The Galactic Halo Ionizing Field}

\author{J. Bland-Hawthorn\\Anglo-Australian Observatory\\P.O. Box
  296\\Epping\\NSW 2121\\
  \\P.R. Maloney\\CASA\\University of Colorado\\Boulder\\CO
  80309-0389} \date{} \maketitle

\begin{abstract}
  There has been much debate in recent decades as to what fraction of
  ionizing photons from star forming regions in the Galactic disk
  escape into the halo. The recent detection of the Magellanic Stream
  in optical line emission at the CTIO 4m and the AAT 3.9m telescopes
  may now provide the strongest evidence that at least some of the
  radiation escapes the disk completely. We present a simple model to
  demonstrate that, while the distance to the Magellanic Stream is
  uncertain, the observed emission measures (${\cal E}_m \approx 0.5 -
  1$ cm$^{-6}$ pc) are most plausibly explained by photoionization due
  to hot, young stars. This model requires that the mean Lyman-limit
  opacity perpendicular to the disk is $\tauLL \approx 3$, and the
  covering fraction of the resolved clouds is close to unity.
  Alternative sources (e.g. shock, halo, LMC or metagalactic
  radiation) contribute negligible ionizing flux.
\end{abstract}

{\bf Keywords:}
interstellar medium $-$ intergalactic medium $-$ individual object: 
Magellanic Stream $-$ Galaxy: corona, halo $-$ interferometry

\section{Introduction}
\nobreak There has been extensive theoretical and observational
interest in establishing what fraction of the total ionizing
luminosity from the stellar disk of the Milky Way and other galaxies
escapes into the halo and the intergalactic medium (\eg Miller \& Cox
1993; Dove \& Shull 1994; Leitherer \& Heckman 1995). Diffuse ionized
gas between HII regions in half a dozen well studied galaxies suggests
that a significant fraction escapes to ionize the ambient ISM (e.g.
Hoopes, Walterbos \& Greenawalt 1996; Ferguson\etal 1996). Broadly
speaking, if the optical depth at the Lyman limit is $\tauLL$, these
observations require $\tauLL\approx 1$ on the scale of the diffuse
disk gas.  The vertically extended Reynolds Layer requires that
$\tauLL \approx 2$ to explain the observed line emission (Reynolds
1990). We now show that the observed \Ha\ emission measures at the
distance of the Magellanic stream (0.5$-$1 cm$^{-6}$ pc in the MS
II$-$IV clumps) are consistent with ionization by the Galactic disk
(Weiner \& Williams 1996; q.v. Bland-Hawthorn 1997), providing $\tauLL
\approx 3$ perpendicular to the disk. More detailed calculations are
given in Bland-Hawthorn \& Maloney (1996).

\section{Galactic photoionization model}
The emission measure ${\cal E}_m$ from the surface of a cloud
embedded in a bath of ionizing radiation gives a direct gauge,
independent of distance, of the ambient radiation field beyond the
Lyman continuum (Lyc) edge (\eg Hogan \& Weymann 1979).  This assumes
that the covering fraction ($\kappa$) {\it seen by the ionizing
  photons} is known and that there are sufficient gas atoms to soak up
the incident ionizing photons.  We assume an electron temperature T$_e
\simeq 10^4$K, as expected for gas photoionized by stellar sources,
for which the Case B hydrogen recombination coefficient is $\aB \simeq
2.6 \times 10^{-13} (10^4/T_e)^{0.75}$ cm$^3$ s$^{-1}$. At these
temperatures, collisional ionization processes are negligible. In this
case, the column recombination rate in equilibrium must equal the
normally incident ionizing photon flux, $\aB n_e N_{H^+} = \varphi_i$,
where \phiI\ is the rate at which Lyc photons arrive at the cloud
surface (photons cm$^{-2}$ s$^{-1}$), $n_e$ is the electron density
and $N_{H^+}$ is the column density of ionized hydrogen. The emission
measure is just ${\cal E}_m = \int n_e n_{H^+}\;dl =n_e n_{H^+} L\ 
{\rm cm^{-6}\; pc}$ where $L$ is the thickness of the ionized region.
The resulting emission measure for an ionizing flux \phiI\ is then
${\cal E}_m = 1.25\times 10^{-2} \varphi_4 \ {\rm cm^{-6}\; pc}$ where
$\varphi_i = 10^4 \varphi_4$.  For an optically thin cloud in an
isotropic radiation field, the solid angle from which radiation is
received is $\Omega = 4\pi$, while for one-sided illumination,
$\Omega=2\pi$.  For the models we will be considering, however,
$J_\nu$ is anisotropic and $\Omega$ can be considerably less than
$2\pi$.

In order to estimate \phiI, we develop an idealized model for
predicting the \Ha\ emission measure at the distance of the Magellanic
Stream.  The ionizing stars are assumed to be isotropic emitters
confined to a thin disk in the $x-y$ plane (or the $X-Y$ plane in
Galactic Coordinates, e.g. Fig. 1).  For a cloud $C$ at position
$(x_0,0,z_0)$ a distance $R$ from an arbitrary patch of the disk $dA$,
the received flux $f_d$ (in units of erg cm$^{-2}$ s$^{-1}$ Hz$^{-1}$)
from ionizing disk sources with specific intensity $\zeta_\nu$ through
a solid angle $d\Omega$ is
\begin{equation}
f_d = \int \zeta_\nu\ {\rm d}\Omega = \int \zeta_\nu\ \cos\theta\ dA(r,\phi) /R^2
\end{equation}
where $dA = r\ dr\ d\phi$ and
\begin{equation}
R = x_0^2 + z_0^2 + r^2 - 2 x_0 r \cos\phi .
\end{equation}
The angle $\theta$ is the polar angle measured from the positive $z$
axis through $dA$ to the line extending from $dA$ to $C$.  Thus, at an
arbitrary point in the galaxy halo, the ionizing photon flux from the
disk (in units of photons cm$^{-2}$ s$^{-1}$) is
\begin{equation}
\varphi_d(r,\phi) = \int n_d(r,\phi)\ \cos\theta\ dA(r,\phi) / R^2 = \int d\sigma(r,\phi)
\cos\theta\ /R^2
\end{equation}
for which $n_d$ and d$\sigma$ are the surface photon density and
brightness, respectively, within each disk element $dA$.

For the opaque disk model, the patch $dA$ is observed through the
intervening disk interstellar medium (ISM) such that $d\sigma^\prime =
e^{-\tauLL(r,\phi)}\ d\sigma$.  For a disk population of OB stars, we
consider an axisymmetric exponential disk with scale length $r_d$,
$n_d(r) = n_0 e^{-r/r_d}$.  We adopt a radial scale length of $r_{d}
=$ 3.5 kpc (Kent, Dame \& Fazio 1991) and all integrations are
performed out to 25 kpc in radius since there is some evidence for
faint HII regions at these large radii (de Geus\etal\ 1993).
Vacca\etal\ (1996) have compiled a list of 429 O stars within 2.5 kpc
of the Sun from which they determine an ionizing surface density of
$n(r_{\odot}) = 3.74\times 10^7$ phot cm$^{-2}$ s$^{-1}$ where
$r_{\odot}$ is the radius of the Solar Circle. After an exhaustive
study of the literature, Reid (1993) finds $r_{\odot} = 8.0\pm 0.5$
kpc.  Thus, from equation (1), we derive $n_0 = 3.7\times 10^8$ phot
cm$^{-2}$ s$^{-1}$.

\section{Photoionization of the Magellanic Stream}

The Stream lies along a great arc which extends for more than
100$^{\circ}$ (\eg Mathewson, Cleary \& Murray 1974).  Fig. 1
illustrates the relationship of the LMC to the Magellanic Stream above
the Galactic disk (Mathewson \& Ford 1984).  We shall make the
assumption that the Stream lies along a circular orbit, close to the
$X$-$Z$ plane, originating from the Lagrangian point between the LMC
and SMC. The Cepheid distance moduli indicate that for the LMC
$(m-M)_0 = 18.47\pm 0.15$, which implies a distance of 49.4$\pm$3.4
kpc (Feast \& Walker 1987); for the SMC, $(m-M)_0 = 18.83\pm 0.15$
which implies 58.3$\pm$4.0 kpc (Feast 1988). Thus, we shall assume an
average galactocentric radius of 55 kpc for the Stream.  This is an
oversimplification since most computed orbits for the LMC-SMC system
imply substantial ellipticity with the Galaxy at a focal point (\eg
Lin, Jones \& Klemola 1995).  Our model is consistent with the
distance measured by Gardiner\etal\ (1994) towards MS VI, but not with
the much smaller value of 20 kpc determined by Moore \& Davis (1994).

In Fig. 2, we present a meridional plot of the halo radiation field
for $\tauLL = 2$.  While the distance to the Magellanic Stream is
uncertain, the expected \Ha\ emission measure for the opaque disk
model should be easily detectable.  For distances of (20,40,60) kpc,
\phiIV\ takes values of (710,215,105) $\times 10^4$ phot cm$^{-2}$
s$^{-1}$ (Fig.  3). From equation (7), the expected ${\cal E}_m$
values are (9.0,2.7,1.3) cm$^{-6}$ pc, or equivalently, (18,5.4,2.6)
$\times 10^{-18}$ erg cm$^{-2}$ s$^{-1}$ arcsec$^{-2}$. The Weiner \&
Williams (1996) detections along the stream are 370, 210 and 200
milliRayleighs\footnote{1 Rayleigh is 10$^6/\pi$ phot cm$^{-2}$
  s$^{-1}$ sr$^{-1}$ or 2.41$\times$10$^{-7}$ erg cm$^{-2}$ s$^{-1}$
  sr$^{-1}$ at \Ha.} or, equivalently, ${\cal E}_m$ values of
(1.1,0.63,0.60) cm$^{-6}$ pc. The \Ha\ measurements of Weiner \&
Williams (1996) are within range of the model values, particularly
since the Stream distance is at the far end of our range.

In Fig. 3, we present the predicted emission measure along the Stream
after projecting the clouds into the $X$-$Z$ plane, where the observer
is assumed to be at the Galactic Centre. If we assume $\kappa$ is
close to unity and remains constant along the Stream, several
conclusions follow immediately. {\it The Galactic disk is unlikely to
  be transparent to ionizing photons otherwise the Magellanic Stream
  would be mostly ionized.} The shape of the \Em\ curve gives an
independent assessment of the disk opacity, but this is sensitive to
departures from a circular trajectory. With relatively few unknowns,
the mean UV opacity of the Galactic disk can be determined after a
comprehensive observational campaign along the Stream.  If the Stream
orbit is highly flattened (Moore \& Davis 1994), the solid line in
Fig. 3 becomes significantly more boxy at large $\delta$, and possibly
even sharply rising towards the edges before turning over. The
expected value of \Em\ at MS VI ($\delta = 135^\circ$) could be almost
an order of magnitude higher for a distance of 20 kpc compared with
our adopted value. The major limitation of our model is the poorly
known cloud geometry and HI covering fraction.

In the interests of brevity, we do not discuss alternative ionizing
sources (e.g. shock or halo sources) as these are expected to be
entirely negligible.  For illustrative purposes only, we include the
expected ionization from the LMC and halo bremsstrahlung in Fig. 3.
For the coronal gas, we assume an isothermal sphere with central
density $2\times 10^{-3}$ cm$^{-3}$, scale length 10 kpc and electron
temperature $2\times 10^6$K (0.2 keV). The LMC is treated as a point
source radiating $5\times 10^{51}$ ionizing photons per second.  For a
complete discussion, we refer readers to Bland-Hawthorn \& Maloney
(1996).

The influence of the corona is only likely to be observable at extreme
$\delta$ angles where emission from the upper cloud face is expected
to dominate.  At $\delta$ angles larger than 150$^\circ$, the
isothermal halo acts much like a distant point source so would be
difficult to distinguish from the LMC ionization.  The LMC radiation
field is not expected to substantially ionize the Magellanic Stream
(MS I$-$VI) although, presumably, it has a major impact on the outer
parts of the Milky Way in the direction $l=270$\deg\ (see Figs. 2 and
3).  If there are no UV-bright companions, the outer extremities of
opaque disks fall inside a `toroidal shadow' which sees only a very
weak ionizing field from the Galactic halo. If the outer warp in the
HI disk is not severe ($\leq 10^\circ$ from center to edge), the
ionization of cold gas at large radius should be dominated by the
cosmic UV background. The current 2$\sigma$ upper limit on the flux,
\phiIV $=3.8$ (q.v.  Bland-Hawthorn 1997), indicates that the cosmic
background is expected to produce an equivalent emission measure less
than \Em\ $= 0.05$ cm$^{-6}$ pc.

In summary, for a mean Stream distance of 55 kpc, if $\kappa=1$, the
\Ha\ detections indicate $\tauLL \approx 3$ perpendicular to the
Galactic disk such that only 5\% of the ionizing radiation from the
disk escapes into the halo.  Notably, Domgorgen \& Mathis (1994) have
obtained the same result using an entirely different approach. While OB
stars should dominate the ionization balance, just how the ionizing
radiation escapes from the star-forming regions into the halo is still
somewhat unclear, although recent theoretical models have begun to
address this issue (Miller \& Cox 1993; Dove \& Shull 1994).

\section*{Acknowledgments} JBH thanks JILA and the University of
Colorado for their hospitality during the early stages of this work,
and Gary da Costa for important references. We gratefully acknowledge
useful comments from J.R. Dickey and an anonymous referee.

\medskip
\reference Bland-Hawthorn, J. 1997, Publ. Astron. Soc. Aust., 14, in press

\reference Bland-Hawthorn, J. \& Maloney, P.R. 1996, ApJ, submitted

\reference Burton, W.B. 1988, In Galactic \& Extragalactic Astronomy,
eds. G.L. Verschuur, K. Kellerman, p. 295

\reference de Geus, E.J., Vogel, S.M., Digel, S.W. \& Gruendl, R.A. 1993, ApJ, 413, 97

\reference Domgorgen, H. \& Mathis, J.S. 1994, ApJ, 428, 647

\reference Dove, J. \& Shull, M. 1994, ApJ, 430, 222

\reference Feast, M.W. 1988, ASP Conf
Ser. 4, 9 (eds. S. van den Bergh \& C. Pritchet)

\reference Feast, M.W. \& Walker, A.R. 1987, ARAA, 25, 345 

\reference Ferguson, A.M.N., Wyse, R.F.G., Gallagher, J.S. \&
Hunter, D.A. 1996, AJ, 111, 2265

\reference Fujimoto, M. \& Sofue, Y. 1976, A\&A, 47, 263

\reference Gardiner, L.T., Sawa, T. \& Fujimoto, M. 1994, MNRAS, 266,
567

\reference Hogan, C.J. \& Weymann, R.J. 1987, MNRAS 225, 1P

\reference Hoopes, C.G., Walterbos, R.A.M. \& Greenawalt, B.E. 1996,
AJ, in press

\reference Kent, S.M., Dame, T.M. \& Fazio, G.G. 1991, ApJ, 378, 131

\reference Leitherer, C. \& Heckman, T.M. 1995, ApJS, 96, 9

\reference Lin, D.N.C., Jones, B.F. \& Klemola, A.R. 1995, ApJ, 439,
652

\reference Mathewson, D.S, Cleary, J.D. \& Murray, M.N. 1974, ApJ,
190, 291

\reference Mathewson, D.S. \& Ford, V.L. 1984, in Structure \& Evolution of the Magellanic Clouds, IAU Symp., 108, 125 

\reference Miller, W.W., \& Cox, D.P. 1993, ApJ, 417, 579

\reference Moore, B., \& Davis, M. 1994, MNRAS, 270, 209

\reference Reid, M. 1993, ARAA, 31, 345

\reference Reynolds, R.J. 1990, in Galactic \& Extragalactic
Background Radiation, eds. S. Bowyer \& C. Leinert, (Dordrecht:
Kluwer), 157

\reference Vacca, W.D., Garmany, C.D. \& Shull, J.M. 1996, ApJ, 460, 914

\reference Weiner, B.J. \& Williams, T.B. 1996, AJ, 111, 1156

\section{Figure Captions}

\nobreak{\bf Figure 1.} An illustration of the LMC and the dominant
clouds in the Magellanic Stream (Mathewson \& Ford 1984). The LMC and
the Stream have been projected onto the Galactic $X$-$Z$ plane. We
have ignored small projection errors resulting from our vantage point
at the Solar Circle. The angle $\delta$ is measured from the negative
$X$ axis towards the negative $Z$ axis where $\delta = -b\ (0^\circ
\leq \delta \leq 90^\circ)$ and $\delta = b+180^\circ\ (90^\circ \leq
\delta \leq 180^\circ)$.  In reality, the orbit of the Stream lies
closer to the Great Circle whose longitude is $l = 285^\circ$.

\nobreak{\bf Figure 2.} Meridional plot showing the probable
contribution of the LMC to the opaque-disk halo radiation field (solid
lines). The dotted lines are for the opaque-disk model in Fig. 3. The
position of the LMC in Galactic coordinates lies within 2 kpc of the
plane $Y=0$ (Fujimoto \& Sofue 1976). The figure shows a 100 kpc
$\times$ 100 kpc intersection of the non-axisymmetric radiation field
in the plane $Y=0$. The dots represent the HI warp in the outer parts
of the Galaxy close to the line of longitude $l=270^\circ$ (Burton
1988).

\nobreak{\bf Figure 3.} The predicted \Ha\ emission measure along the
Magellanic Stream as a function of $\delta$. The vertical
axis has units of log(cm$^{-6}$ pc); these can be converted to
log(Rayleighs) by subtracting 0.48.  The dotted curves (top) assume an
optically thin Galactic disk with and without the LMC ionizing field.
The solid lines assume an opaque ionizing disk with (thin line) and
without (thick line) a bremsstrahlung halo. The LMC contribution to
the opaque disk($+$halo) model is shown by the short-dash ($\tauLL =
2$) and long-dash ($\tauLL = 2.8$) curves.  The dot-dash curve is
\Em(\Ha) predicted from the upper side of the Magellanic Stream due to
the bremsstrahlung halo. The solid points are the \Em\ measurements of
Weiner \& Williams (1996).

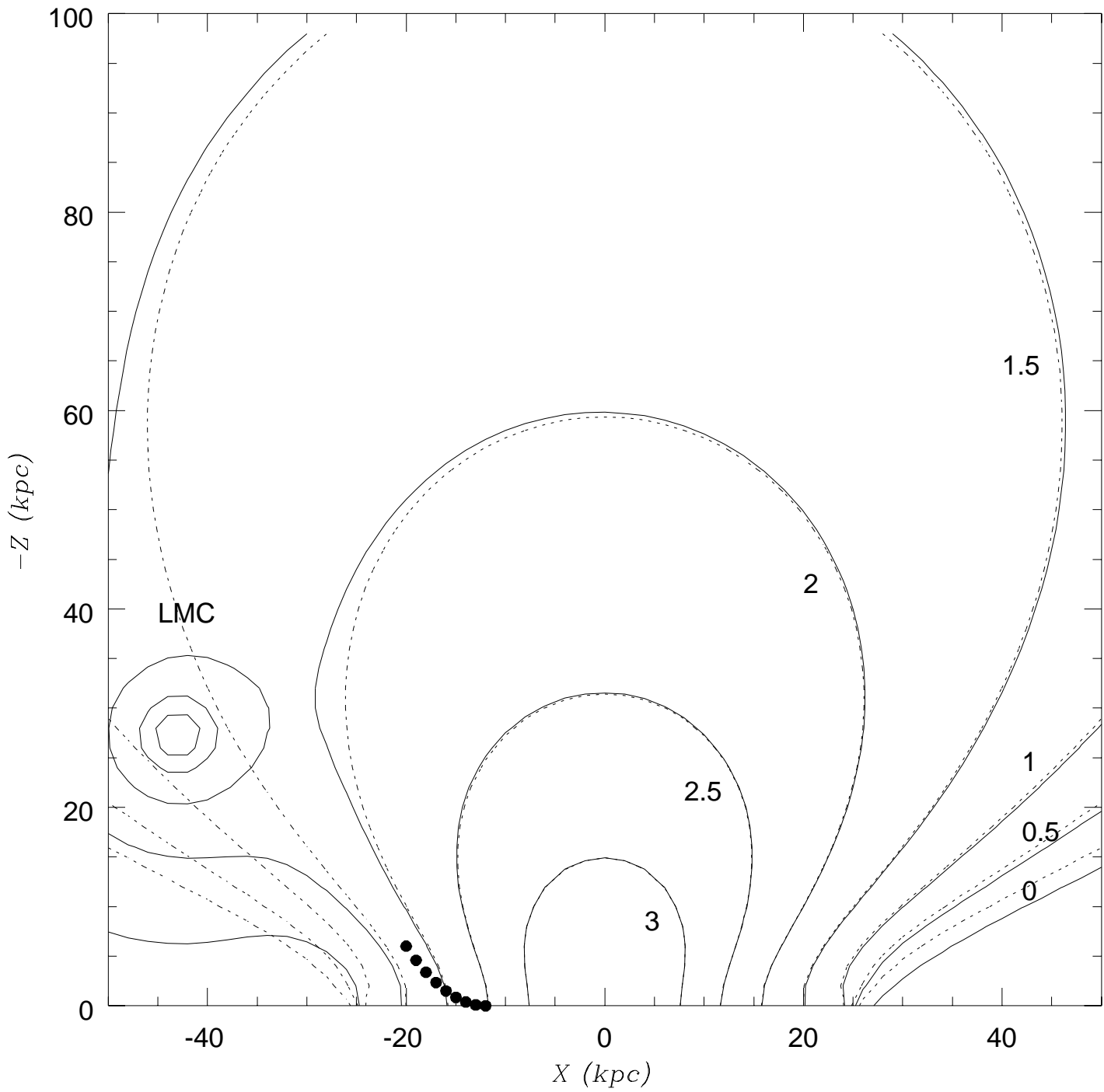
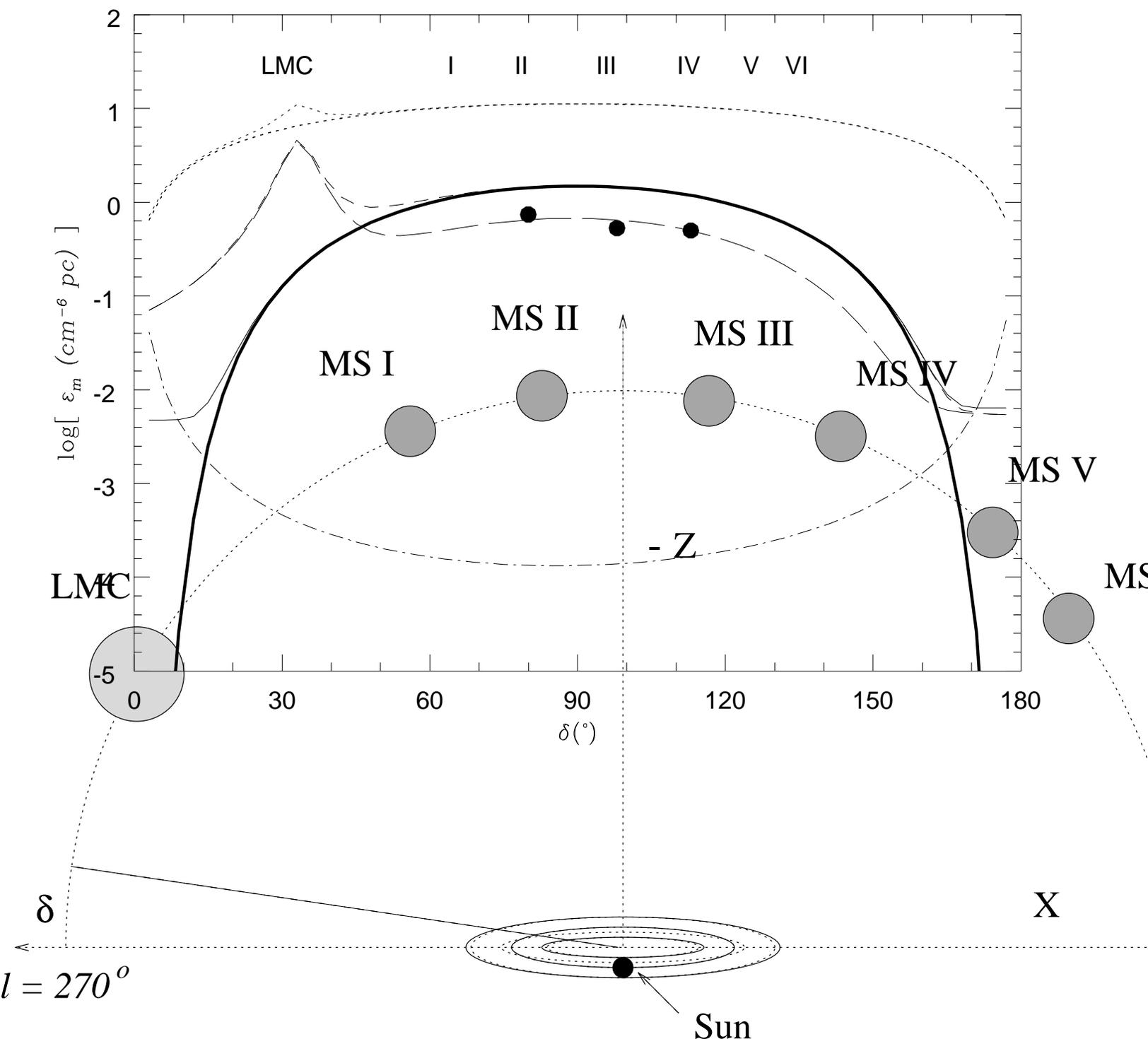
\end{document}